\documentclass[fleqn,usenatbib,useAMS]{mnras}

\usepackage{graphicx}	
\usepackage{amsmath}	
\usepackage{amssymb}	
\usepackage{textcomp}   
\usepackage{multicol}   
\usepackage{bm}		    
\usepackage{pdflscape}	
\usepackage{natbib}	

\usepackage[T1]{fontenc}
\usepackage{ae,aecompl}
\usepackage[latin1]{inputenc}
\usepackage{multirow}
\usepackage{setspace}

\defcitealias{Devaraj2018POLICAN}{Devaraj2018a}
\defcitealias{Devaraj2018CANICA}{Devaraj2018b}

\title[The Near-Infrared Polarization of Frosty Leo]{The Near-Infrared Polarization of the Pre-Planetary Nebula Frosty Leo}
\author[E.O. Serrano Bernal ]{
\noindent E.O. Serrano Bernal$^{1}$,\thanks{E-mail: arthas@inaoep.mx}
L. Sabin$^{2}$, A. Luna$^{1}$, R. Devaraj$^{1,3}$, Y. D. Mayya$^{1}$ and \newauthor 
\noindent L. Carrasco$^{1}$
\\
$^{1}$Instituto Nacional de Astrof\'isica, \'Optica y Electr\'onica, C.P. 72840, San Andr\'es Cholula, Puebla, M\'exico.\\
$^{2}$Instituto de Astronom\'ia, Universidad Nacional Aut\'onoma de M\'exico, Apdo. Postal 877, 22860 Ensenada, B. C., M\'exico.\\
$^{3}$Dublin Institute for Advanced Studies, Astronomy and Astrophysics Section, 31 Fitzwilliam Place, Dubin 2, Ireland\\
}
\date{Accepted XXX. Received YYY; in original form ZZZ}
\pubyear{2020}
\begin{document}
\label{firstpage}
\pagerange{\pageref{firstpage}--\pageref{lastpage}}
\maketitle
%
\begin{abstract}
We present a near-infrared imaging polarimetric study of the pre-planetary nebula: Frosty Leo. 
The observations were carried out in J, H and K' bands using the new polarimeter POLICAN mounted on the 2.1m 
telescope of the Guillermo Haro Astrophysical Observatory, Sonora, Mexico. The most prominent result observed
in the polarization maps is a large and well defined dusty envelope (35\arcsec\ diameter in H-band).
The polarization position angles in the envelope are particularly well ordered and nearly parallel 
to the equator of the nebula (seen in J and H bands).
The nebula presents a known bipolar outflow and the envelope completely wraps around it.
Within the bipolar lobes, we find high polarization levels ranging from $60\%$ (J band) to $90\%$
(K' band) and the polarization angles trace a centrosymmetric pattern.
We found the remnants of superwind shells at the edges of the bipolar lobes and the duration of this phase is around 600 yrs.
The origin of polarization features in the nebula is most likely due to a combination of single and multiple scattering.  
Our results clearly demonstrate new structures that provide new hints on the evolution of Frosty Leo from its previous asymptotic giant branch phase.   
\end{abstract}



\section{Introduction}
Pre-planetary nebulae (PPNe) are evolved low-intermediate mass stars in a transitional stage between 
the asymptotic giant branch (AGB) and planetary nebula phases \citep{Kwok1993}. 
PPNe display a circumstellar envelope (CSE) of dust and gas which was formed as a result of 
the mass-loss (up to $10^{-5} M_\odot yr^{-1}$ \citealt{Loup1993}) suffered during their previous AGB phase. 
The CSEs keep records of the mass-loss history and in consequence, 
their study can help to understand their, often, intriguing morphology.\\
PPNe are mostly seen by scattered light in the optical and near-infrared (NIR), so, they are expected to be linearly polarized.
Taking advantage of this, \citet{Gledhill2001,Gledhill2005} performed 
NIR imaging polarimetry not only to detect the faint CSEs from several PPNe, 
but also to resolve structures within them. 
Thus, polarimetry is a useful tool to study the CSE of PPNe.\\ 
In this work, we present
the results of the polarimetric observations of the well-known pre-planetary nebula Frosty Leo (IRAS 09371+1212). 
It was first identified as a post-AGB with a bipolar structure by \citet{Forveille1987}. 
Later, NIR images and spectroscopic observations confirmed its morphology and the presence of ice water 
\citep{Hodapp1988}. By assuming that the central star has a spectral type K7III or II, \citet{Mauron1989} estimated 
a distance between 1.7 kpc and 4.3 kpc, and the stellar temperature which corresponds to the 
spectral type is 3750 K \citep{Robinson1992}.\\ 
Several works investigated this intriguing object 
and due to its dusty nature, polarimetry appeared as a useful tool. 
By means of imaging polarimetry, \citet{Clemens1987}
unveiled an extended but irregular polarized envelope at visible wavelengths. 
In the NIR, polarimetric studies of Frosty Leo have focused on its central region. 
The first analysis was performed by \citet{Dougados1990} using the 3.6m CFHT and the CIRCUS IR camera.
Within a  small region of $8\arcsec \times\ 8\arcsec$, 
they detected a centrosymmetric polarization pattern as well as 
the signature of an edge-on warped dust disk. Later on,  \citet{Murakawa2008} 
used the 8m Subaru telescope equiped with a coronagraph and adaptive optics to observe the central part
of Frosty Leo within a $10\arcsec \times\ 10\arcsec$ area. 
They reported some new polarimetric features such as an almost straight depolarized region
and a slight vector alignment in a small spot on the equator.\\
High resolution imaging polarimetry has allowed a better understanding of Frosty Leo's central part. 
However, a lot can still be learned from this object at a larger scale. Indeed, 
a bigger Field-of-View (FoV) is desirable when it comes to the study of the outer structure. 
For this object, imaging polarimetry was performed with the NIR polarimeter 
of Cananea, POLICAN \citep{Devaraj2018POLICAN}. The combination
of a large FoV and an instrument with good sensitivity, even with a 2m telescope, can lead to some new findings as it will be presented 
in the sections ahead. The structure of the paper is as follows: we describe the observations and data reduction in section 2,
in section 3 we present our results, discussion of the observed polarization will be carried on in section 4 
and our concluding remarks are shown in section 5.\\
\section{Observations and Data Reduction}
Imaging polarimetry of Frosty Leo was performed at the 2.1m 
Guillermo Haro Astrophysical Observatory (OAGH) telescope, 
at Sonora, Mexico, using POLICAN mounted on the Cananea near infrared camera (CANICA) \citep{Carrasco2017}.
The $1024 \times\ 1024$ element HgCdTe detector is sensitive between 0.85 and 2.4 microns,
and was used in conjunction with J($1.24\mu$m), H($1.63\mu$m) and K'($2.12\mu$m) broad-band filters.
The 2:1 focal reducer of CANICA provides a scale plate of 0.32\arcsec\ per pixel and the FoV of around 5\arcmin.\\
The single-beam polarimetric module of (POLICAN) consists of a rotating half-wave plate (HWP) and a wire grid linear polarizer 
placed in front of CANICA. Both elements are placed near the focal plane and their diameters 
allow to use the full 5\arcmin\ FoV. In order to measure the linear polarization, the HWP 
is rotated to 4 position angles ($0^{\circ},45^{\circ},22.5^{\circ},67.5^{\circ}$) to complete 
1 set of four images. Several sets of dithered images are obtained to calculate and subtract the sky emission to
later compute the Stokes Parameters. 
Major insight of POLICAN's features and the standard observing methods are described
in detail by \citet{Devaraj2018POLICAN}.\\
The observations were carried out during 2 runs. The first took place during April 14 and 15 of 2017 and the 
second one was carried out during the nights of February 23 to 25 of 2018. Weather conditions 
were good for both runs and the seeing was about 1\arcsec. For J and H bands we used the standard "dithering scheme", in which
10 dithered sets of images were obtained with exposure time of 30 seconds per frame.
This results in a total integration time of 1200 secs.
For K' band we followed a similar scheme, although, to avoid background saturation and boost signal-to-noise ratio,  
the images were co-added 3 times with a exposure time of 20 seconds.
The total integration time in K' band was 2400 seconds (see table \ref{tab:t_y_mags}).\\
Data reduction for POLICAN consists of 3 steps: first, standard corrections as 
flat fielding, dark current and background subtraction are performed. 
Later, the images are stacked and aligned to a common coordinate system,  
several {\sc IRAF} packages were used to develop such routines.
In the second step, the Stokes parameters $Q$ and $U$ are calculated and later corrected
to eliminate any instrumental polarization, such as the HWP offset angle, $\theta_{off}$, and the intrinsic degree of polarization $P_{inst}$.
Finally, the polarization angle, $PA=0.5\times\ tan^{-1}(U/Q)$, polarization degree, $P\%=100\sqrt{U^2 + Q^2}$, 
total intensity $I$, and error maps are computed using {\sc Python} software packages.  
All maps presented here have a binning of $3 \times 3$ pixels to match the 1\arcsec\ seeing.
The maximum errors in P$\%$ and $PA$ are 12$\%$ and $10^{\circ}$, respectively.  
The NIR magnitudes from the 2MASS catalogue \citep{Cutri2003} and the integration
times are shown in table \ref{tab:t_y_mags}.\\
\begin{table}
\caption{NIR magnitudes (2MASS) of Frosty Leo and integration time for each band.}
\label{tab:t_y_mags}
\begin{tabular}{@{\extracolsep{45pt}}|l|l|l|}
\hline
POLICAN                & 2MASS        &  $t_{int}$ $(s)$ \\
IR Band                & magnitude    &                  \\ \hline                                  
J ($1.24\mu m$)   & 8.06              &   1200            \\ 
H ($1.63\mu m$)   & 7.48              &   1200            \\ 
K'($2.12\mu m$)   & 7.46  (K)         &   2400            \\ 
\hline
\end{tabular}
\end{table}
\begin{figure*}
	\includegraphics[width=\textwidth]{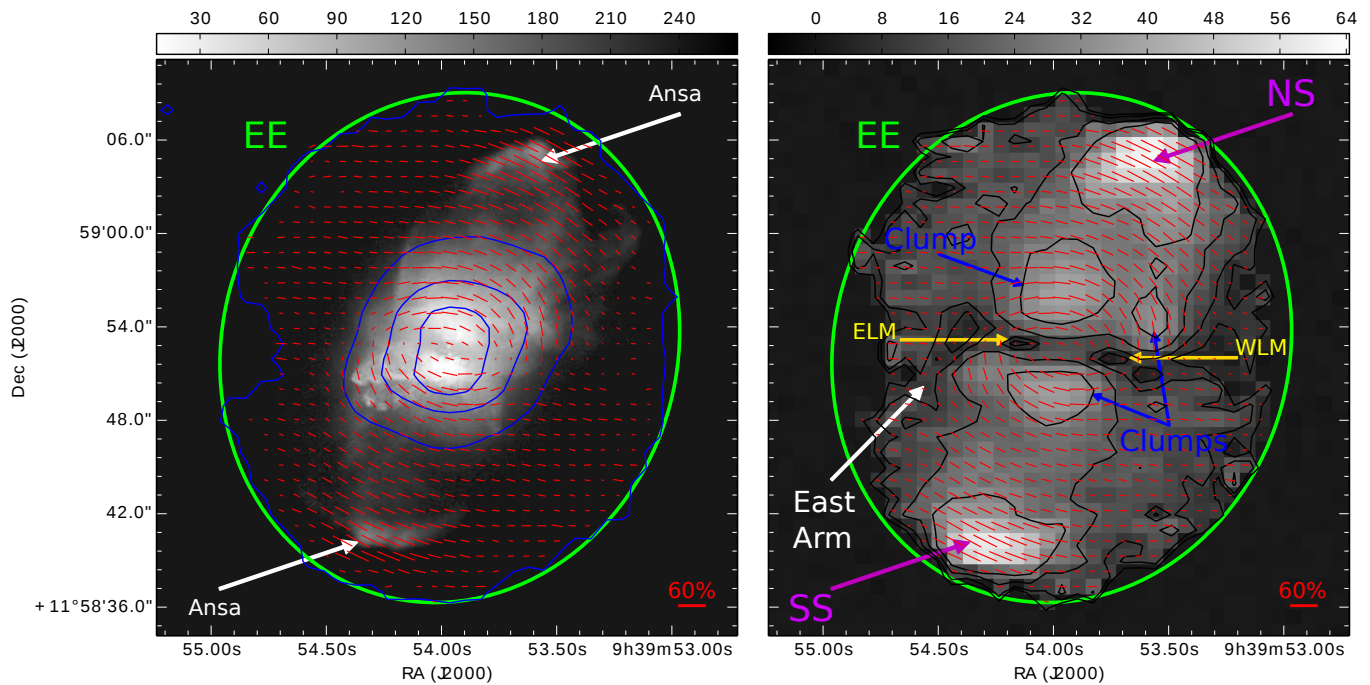}
    \caption{Polarization maps of Frosty Leo in J band.
      The left panel shows an HST composite image (F606W and F814W filters) of Frosty Leo 
      superimposed with (red) electric vectors. The total intensity is displayed every 2 magnitudes in blue contour levels (arbitrary units).
      The right panel shows the degree of polarization ($P\%$) as gray-scale image.
      Selected contour levels ($8\%$, $17\%$, $20\%$, $30\%$)  
      are displayed to highlight the structures unvealed by our data. 
      The regions of interest are labeled according to text. In both panels, an ellipse of $33$\arcsec\ $\times$ $29$\arcsec\ is displayed
      to indicate size and geometry of the extended envelope and the scale is shown at the top.
      }
    \label{fig:J}
\end{figure*}
\section{Results}
\label{sec:Results}
Figures \ref{fig:J}, \ref{fig:H} and \ref{fig:K} show the J, H and K' bands polarization maps for Frosty Leo, respectively. In
all 3 figures, the intensity (I Stokes) and polarization degree ($P\%$) are shown with vectors superimposed. 
The vectors are parallel to the electric field E and their length is proportional to the degree of linear polarization, a 
reference scale vector is also shown in each map.\\ 
We emphasize that the maps in figures \ref{fig:J} and \ref{fig:H} display a noticeable sinking to decrease in intensity in the eastern region.
This might be misinterpreted as a waist-like feature, but it is an instrumental artifact
caused by cross-talk between pixels during readout.
Our reduction pipeline is able to remove this undesirable effect to some degree but not completely \citet{Devaraj2018CANICA}. Nevertheless,
the data presented here are still reliable.\\ 
\subsection{Polarization in J band}
The intensity emission (fig. \ref{fig:J}, left panel) reaches its peak at the central star location and 
then it decreases almost evenly in all directions, except at tips of the bipolar lobes, where a small increase is detected.\\
The NIR polarization in Frosty Leo was previously studied and reported by \citet{Dougados1990} in J band, although their research 
was only focused whithin a radius of ~5\arcsec.
They found $P\%$ as high as ~$40\%$ in the lobes close to the central star and a low polarized region across the equator.
Our polarization map is in good agreement for the zones in common and extend this study to 3 times more radius.\\
The first noticeable feature in our maps is the extended envelope (EE) that protrudes from the well-known bipolar lobes of this object. Part of the EE is also seen at visible wavelengths,
\citet{Clemens1987} detected it also employing imaging polarimetry. 
However, our new POLICAN data reveal a larger size.
To our knowledge, it is the \textit{first detection} of this extended feature in the NIR.
Since this object is fainter in the J band, the shape of the EE looks elliptical at this wavelength.
However, based on our data from the H band we conclude that the actual shape is more likely spherical (see fig. \ref{fig:H}).  
We did not see any variation of the $P\%$ within the EE and we estimated an average of $16\%$. 
The polarization vectors related to this EE show a clear east-west alignment, and the average
PA of these vectors is $79^{\circ}\pm 10^{\circ}$.\\
Whitin the bipolar outflows, the polarization vectors follow a centrosymmetric pattern. 
Such vector behavior suggests single scattering as the polarizing mechanism and that 
the optical depth in the media is low \citep{Gledhill2001,Gledhill2005}.
Previous observations at visible wavelengths obtained similar results \citep{Scarrott1994}.
Within a radius of 7\arcsec\ we resolved 3 clumps (2 northwards and 1 southwards from equator),
these structures are characterized by their $P\%$ which are between $30\%$ to $40\%$.
A narrow and relatively low polarized area is seen between the north clumps. It could be identified with 
the elongated region found by \citet{Murakawa2008} in H and K bands. 
The authors attributed the low polarization level to a low dust density, which is in good agreement with our interpretation.
Another two clumps are found at the tips of the bipolar jets and because of their appearance, 
we labeled them NS and SS as in north shell and south shell, respectively. 
They seem to match the location of the ansae remarked by \citet{Sahai2000} 
and their $P\%$ ranges between $40\%$ and $62\%$.\\
Along the equator, there is a narrow region of low polarization, often associated with a disk 
\citep{Rounan1988,Hodapp1988,Dougados1990,Murakawa2008}. 
Some of the vectors in this region are aligned and parallel to the equator, these vectors are localized between 
2 spots of almost zero polarization:  
the east local minima (ELM) and the west local minima (WLM) (see fig. \ref{fig:J} right panel).
\citet{Murakawa2008} found similar results in H and K bands 
and also suggested the dichroic mechanism as a possible explanation.
A dichroic interpretation would indicate the presence of non-spherical grains with their
longer axes preferentially aligned. The electric vector in the transmited beam is thus 
perpendicular to that preferential direction \citep{lazarian2003,whittet2004,Lazarian2007,Crutcher2012}.\\
On the eastern edge of the equator,
we notice an "arm-like" region of 7\arcsec\ length that is highlighted by its low polarization ($11\%$)
and is directed towards the south.
A low polarization trend is also seen towards the northwest, but it looks more diffuse and
rapidly reaches our detection limit. \citet{Sahai2000} labeled a structure as northwest lobe in that same direction.
\subsection{Polarization in H band}
\label{sec:H}
\begin{figure*}
	\includegraphics[width=\textwidth]{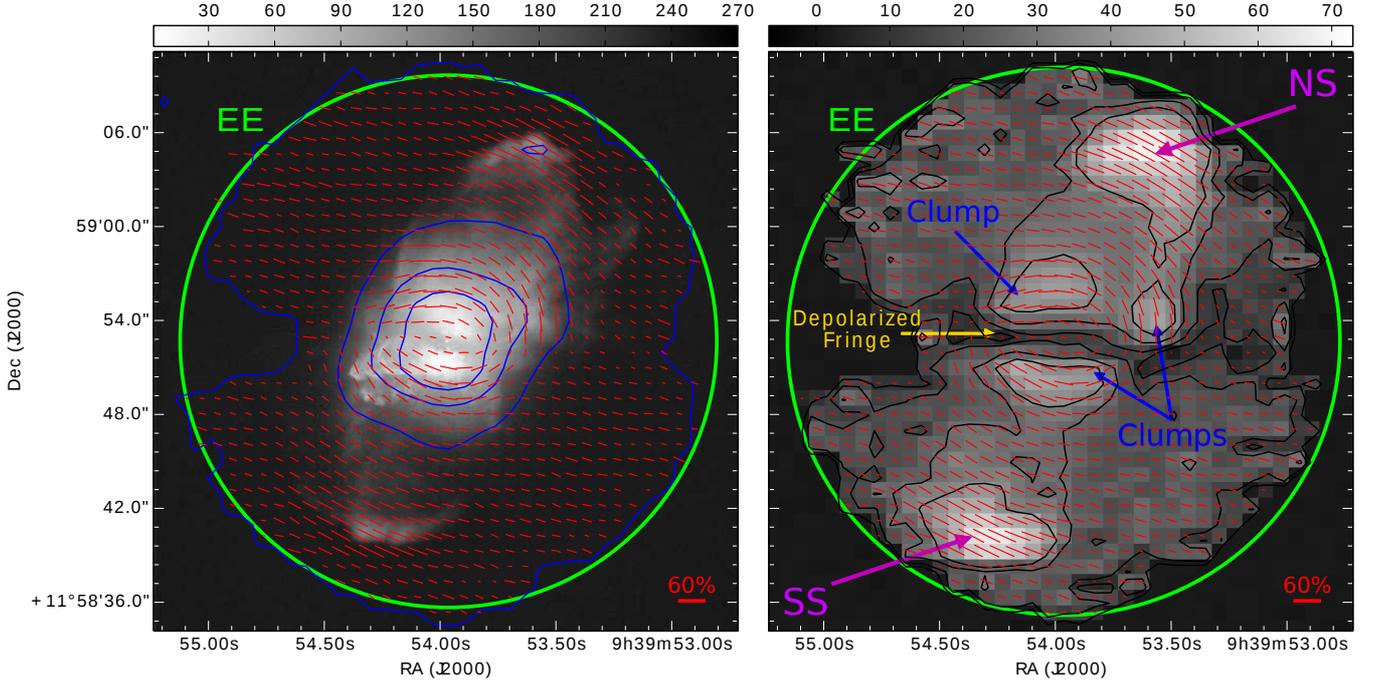}
    \caption{Polarization maps of Frosty Leo in H band.
     Details are as for figure \ref{fig:J}. 
     Contour levels in the right panel correspond to $16\%$, $26\%$ and $36\%$ of polarization.
     The circle in both panels has a diameter of 35\arcsec.}
    \label{fig:H}
\end{figure*}

The polarization maps in H band are shown in fig. \ref{fig:H}.
The total intensity of Frosty Leo in H band follows a trend quite similar to that in J band. 
Except for the low polarized arm-like area and the equatorial local minima, the polarization in H band looks quite 
similar to that in the previous section.
We emphasize that polarization features such as the EE and the north and south shells are first time detections in the NIR.\\
Although our resolution is only $1$\arcsec, our results are 
in good agreement with those shown by \citet{Murakawa2008} (adaptive optics psf of $0.16\arcsec \times 0.28\arcsec$ in K band). 
Whithin a radius of $7$\arcsec, the $P\%$ reaches between $40\%$ and $50\%$ in the clumps nearby the central star, and a
depolarized fringe along the equator shows an average of ~$11\%$. As in J band, some vectors 
in this fringe are parallel to those in the EE. This depolarized region seems to extend towards the east and west almost
as if splitting the whole nebula in half.\\
Although slightly distorted by the eastern depression, the EE in H band seems mostly spherical and considerably larger than in the J band image.
Such difference in size is probably a matter of detection limit since Frosty Leo is brighter in H band by almost one magnitude
(see table \ref{tab:t_y_mags}). 
The diameter of the EE is of 35\arcsec\ and the vectors linked to this area are clearly aligned on east-west direction. 
We measured an average PA is $74^{\circ}\pm~10^{\circ}$ 
and the $P\%$ has a mean value of $20\%$, which is slightly larger than in J band (see table \ref{tab:resumen}).\\
Along the major axis, the vectors follow a centrosymmetric pattern and 
the NS and SS are clearly seen in polarized light. The $P\%$ ranges from $36\%$ to $70\%$ 
and their appearance is more spreaded out than in J band. The NS  
seems to smear down southwards, almost reaching the internal clumps.\\
\begin{figure*}
	\includegraphics[width=\textwidth]{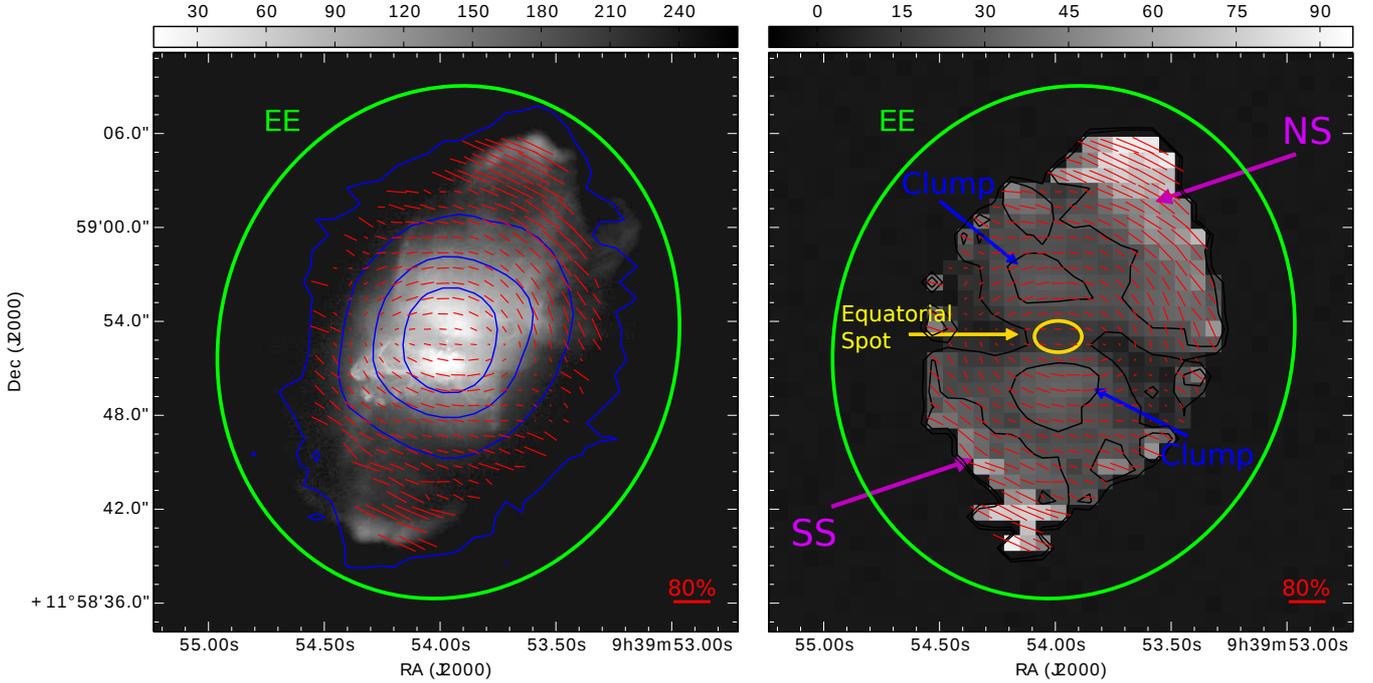}
    \caption{Polarization maps of Frosty Leo in K' band.
      Details are as for figure \ref{fig:J}.
      The contour levels in the right panel correspond to 13$\%$ and 25$\%$ of polarization.
      In both panels, an ellipse of $33$\arcsec\ $\times$ $29$\arcsec\ is displayed for comparison.}
    \label{fig:K}
\end{figure*}
\subsection{Polarization in K' band}
\label{sec:K}
K' band polarization map of Frosty Leo is shown in figure \ref{fig:K}. 
This is (to our knowledge) the first polarization map presented in this band for this object. 
Given our detection limit, the EE is not seen in this map and only the bipolar nebula was clearly detected.
Nevertheless, it still provide some new information on this object.   
As expected, most of the vectors in the map follow a centrosymmetric pattern. The exception is a small equatorial spot (ES) 
where several vectors are perpendicular to the nebula's axis.
Although not entirely, the ES seems to match the location of some aligned vectors 
in the J and H images, this is better seen in fig. \ref{fig:aligned}.
The vectors in the ES show an almost constant position angle of $80^{\circ}$ whilst the average degree of polarization is $15\%$.\\  
The polarized structure in this band only shows two clumps near the equator.
The "missing clump" seems to be merged with the NS. This could be an effect of poor resolution since the PSF 
in this band is degraded to ${\sim}$1.5\arcsec. 
The polarization in the central clumps range from $25\%$ and $35\%$.\\
As in the previous maps, the highest $P\%$ levels are
found in the NS and SS, in some points even reaching or exceeding the $80\%$. These high 
P$\%$ levels are not uncommon since other PPNe such as OH 231.8+4.2 shows similar levels 
and this is in agreement with a scattering process \citep{Ageorges2000}. 
The NS looks more extended and is also smearing southwards even more evidently than in H band.
The SS is less evident given the limits of our map. Nevertheless, a large amount of material is 
seen at the edge of the southern side of the map and it could still be associated to the SS.
\section{Discussion}
\subsection{The polarization in the NS and SS of Frosty Leo}
\begin{figure*}
        \includegraphics[width=15cm]{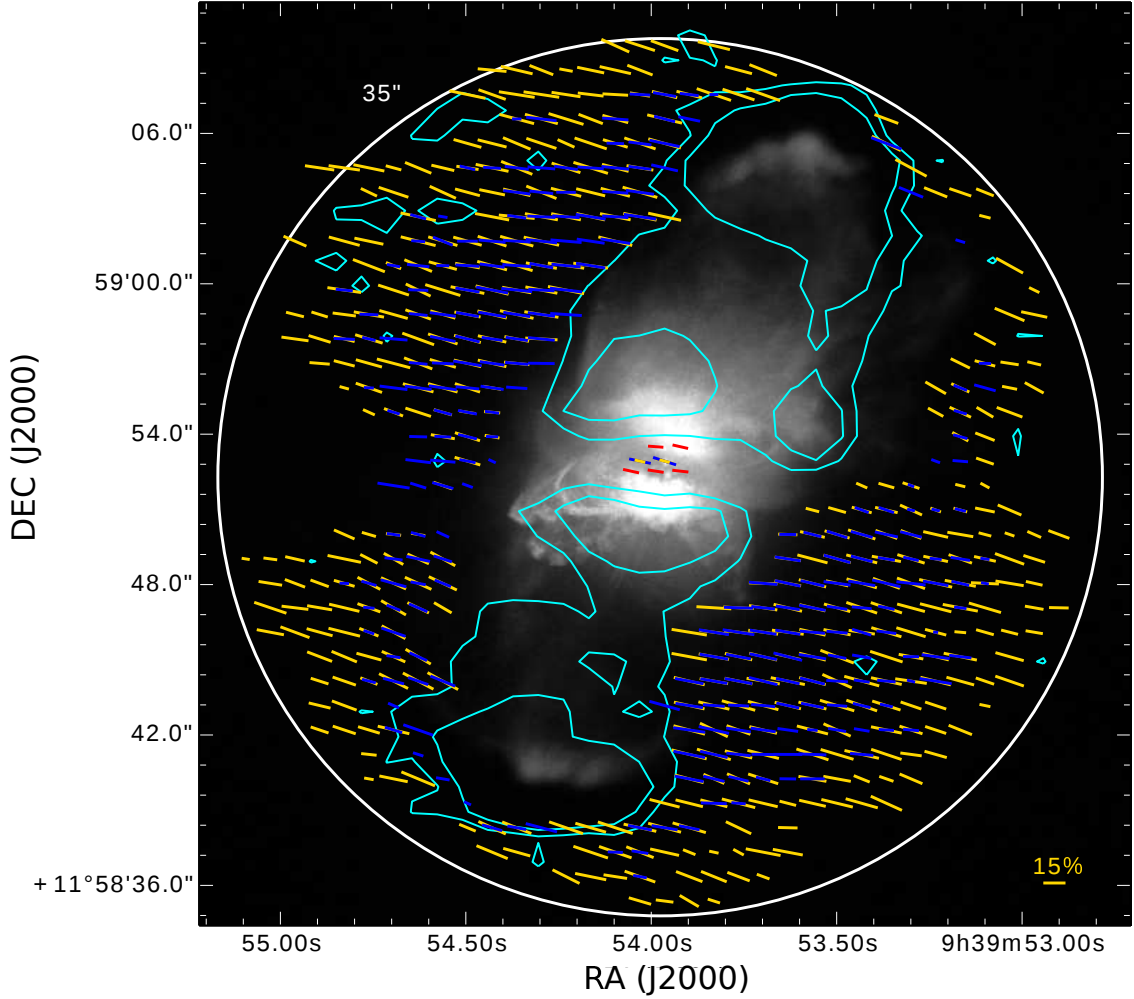}
    \caption{HST composite images (F606W and F814W filters) of Frosty Leo superimposed with polarization vectors.
      Polarization in the J (blue), H (yellow) and K' (red) bands are displayed.  
      The vectors that follow a centrosymmetric pattern have been removed in all 3 bands. 
      Green contour levels correspond to polarization in H band (same as in fig. \ref{fig:H}) 
      and are displayed to remark the inner clumps and the north and south shells.
      The white circle delineates the EE which has a diameter of 35\arcsec\ in H band.
      Similar PAs are observed between bands in both, the extended emission and the equatorial spot.}
    \label{fig:aligned}
\end{figure*}

AGB stars produce a CSE through a slow and tenuous wind \citep{Kwok1993},
the moderate rate of mass ejection through the AGB-wind is about $10^{-5}M_{\odot}yr^{-1}$ \citep{Loup1993}. 
At the end of the AGB lifetime, the mass-loss rate and the velocity of the expelled material
increase drastically (up to ${\sim}10^{-4}M_{\odot}yr^{-1}$ e.g. \citealt{Bujarrabal1999,Bujarrabal2001}).
This period of accelerated mass-loss is expected to last a few thousand years \citep{Kaufl1993} and is known as the superwind phase.\\
The global shaping of the nebula must occur at this stage of increased mass-loss \citep{Lagadec2016}.
A noteworthy feature seen in many nebulae are the ansae. 
Ansae are double knot-like structures along the nebula's axis \citep{Aller1941}.
These knots move in opposite directions and often, their speed is faster than the previously AGB ejected material 
(e.g. \citealt{Patriarchi1991,Redman2000,Sahai2015}).
The mechanism that leads to ansae formation is still unclear, but,
since they are oriented along the major axis, it is probable that they are
the remnants of outflow collimation.
The most likely sources of collimation are the mass-transfer between a binary system (e.g. \citealt{Soker2000,Jones2017}),
or a magnetic field (also driven by a binary companion) \citep{Nordhaus2006,Nordhaus2007}.
Once collimation has taken place, 
ansae could be formed by diverse mechanisms (see \citealt{Balick2002} for a discussion of ansae formation).\\
Although their origin is not fully understood, ansae (or shells) can be used to constrain the superwind time $t_{sw}$ \citep{Gledhill2001}.
The outer and inner edges of a shell should indicate the beginning and cessation of
mass-loss. Thus, by using the shell's angular size and assuming expansion velocities from molecular line measurements,
we can estimate the superwind time \citep{Meixner1997,Gledhill2001}.\\
The bipolar regions within the nebula are optically thin (centrosymmetric pattern),
thus, the NS and SS are physical structures illuminated by a central source \citep{Wolstencroft1986,Dougados1990,Gledhill2001}.
Since these shells are symmetric, detached from the central region and spatially correlated with ansae (see fig. \ref{fig:aligned}), 
we are assuming that they are the remains of a superwind event.\\
Our polarization maps show that the NS and SS cover an area of 51 and 42 $arcsec^{2}$, respectively. By inspection of
the J band map, where the resolution is better, the average angular size of the shells (along the polar axis) was found to be $8\arcsec$.
At a distance of 3 $kpc$ \citep{Mauron1989,Robinson1992}, their length would be of ${\sim}0.1~pc$.
To estimate the superwind time $t_{sw}$, we used the CO speed component of 50 $km~s^{-1}$ measured by \citet{Sahai2000} 
and an inclination angle of $15^{\circ}$ \citep{Roddier1995}:
\begin{equation*} 
t_{sw} \simeq \frac{l_{sh}}{V_{exp}} \simeq \frac{(3~kpc)\tan (8\arcsec)}{(50~km~s^{-1})/(\sin 15^{\circ})} \simeq 600~yr
\end{equation*}.\\ 
where, $l_{sh}$ is the length of the shell and $V_{exp}$ is the projected expansion velocity of the CO along the axis. 
Given the assumption of a constant speed when it actually may vary with time, this estimation should be considered as a lower limit. 
The computed error is $200~yr$ and it is mostly due to the assumed distance to the object.
Nevertheless, this estimation is not unrealistic when compared with the expected duration 
of this phase ($t_{sw}$ ~ $10^3-10^4$ $yr$ \citep{Kaufl1993}).\\
If a binary is involved in the shaping of the nebula, a common envelope evolution could provide
enough rotation to start a dynamo effect. This dynamo would be able to 
produce an explosive outflow from which ansae could be created \citep{Nordhaus2006,Nordhaus2007}.
Considering this, \citet{Nordhaus2006} derived a burst time of ${\sim}100~yr$ for the ansae in the planetary nebula NGC 7009.
This time seems short when is compared with the $600~yr$ that we calculated in the shells of Frosty Leo. However, 
since Frosty Leo is a younger object and its distance is somewhat uncertain, it might be worth it to investigate if   
a dynamo could have produced the ansae with further modelling.\\
A computational model that could fit better with our results was proposed by \citet{Steffen2001}. 
The so-called "stagnation knots" model requires a time-dependent jet that drives into a CSE forming knot-like structures (shells).
At some point, the stream of material will stop and the shell slows down as it expands
and continues adding mass from the envelope.
This model can reproduce morphologies and speeds ($240~kms^{-1}$) similar to those we inferred in the NS and SS 
for an assumed distance of 3 kpc (see model b from \citealt{Steffen2001}), 
although, the time required to form shells is 
a bit larger (${\sim}1100~yr$) than our $600~yr$ calculation. However, if we regard that our estimation is a lower limit, 
both ages are still comparable.\\
\subsection{The polarization of the Extended Envelope in Frosty Leo}
The extended envelope in Frosty Leo and its geometry are seen in J and H bands (figs. \ref{fig:J} and \ref{fig:H}). 
The largest angular size of the faint emission was detected in the H band where it shows a diameter of 35\arcsec. 
This EE can be understood as the remains of the previous AGB phase ejecta. 
Since the halo's shape seems spherical, the mass-loss in the first stages was probably isotropic.
The polarization from J and H bands did not reveal new structures or trends.
Nevertheless, the most noteworthy feature 
is the well-defined alignment from most of its polarization vectors (79$^{\circ}$ and 74$^{\circ}$, in J and H bands, respectively).\\
\citet{Gledhill2005} proposed that the vector alignment is caused by an optically thick media,
under this assumption, most of the material in the nebula
should concentrate at the equator in a dense disk. The disk is then responsible for the high optical depth 
and the alignment distribution. However, a high optical depth
would have prevented us from detecting any structure in the bipolar regions, thus, this interpretation does not stand for the EE.\\
Since the EE (35\arcsec) is larger than the resolution of our data (1\arcsec),
the PSF smoothing effect is also an unlikely answer \citep{Piirola1992,Gledhill2005}.\\  
\citet{Whitney2002} showed that single scattering could also produce non-symmetric polarization, for this,
the grains should be aligned by a magnetic field. In this model,
a single source illuminating an optically thin cloud of non-spherical grains can account for some asymmetries in
the polarization vector pattern. This could fit with our observational results assuming that the EE is 
illuminated directly by the star but the grains need to be coherently aligned by a strong magnetic field.
Observations of OH Zeeman splitting and submillimeter polarimetry have shown magnetic fields of the order of mG in 
post-AGB stars \citep{Bains2003,Sabin2007}. However, for this object, more observational evidence is required.\\ 
Multiple scattering is another alternative. This could occur when an envelope is not 
directly illuminated by the star, but by the light that has been already scattered. Previously scattered light reaching the envelope, 
could be scattered again by the material within and this results in an aligned
polarization pattern \citep{Bastien1988,Fischer1994,Scarrott1995,Wolf2002}.
A scenario where multiple scattering could account for our observational results is the following: 
The bipolar lobes are directly illuminated by the central star and scatter the light as a first event. 
Then, this light later reaches the EE and following scattering events could occur and result in the aligned polarization pattern.
This scenario is quite similar to that proposed by \citet{Bastien1988}, only in their case, the subsequent scattering events occur
in a circumstellar disk.\\
We favor an interpretation in which the EE is illuminated by scattered light coming from the lobes. 
This is the more likely physical situation for the EE since the scattering dominates in the NIR \citep{Scarrott1991,Bains2003}.\\
%
\begin{figure}
	\includegraphics[width=\columnwidth]{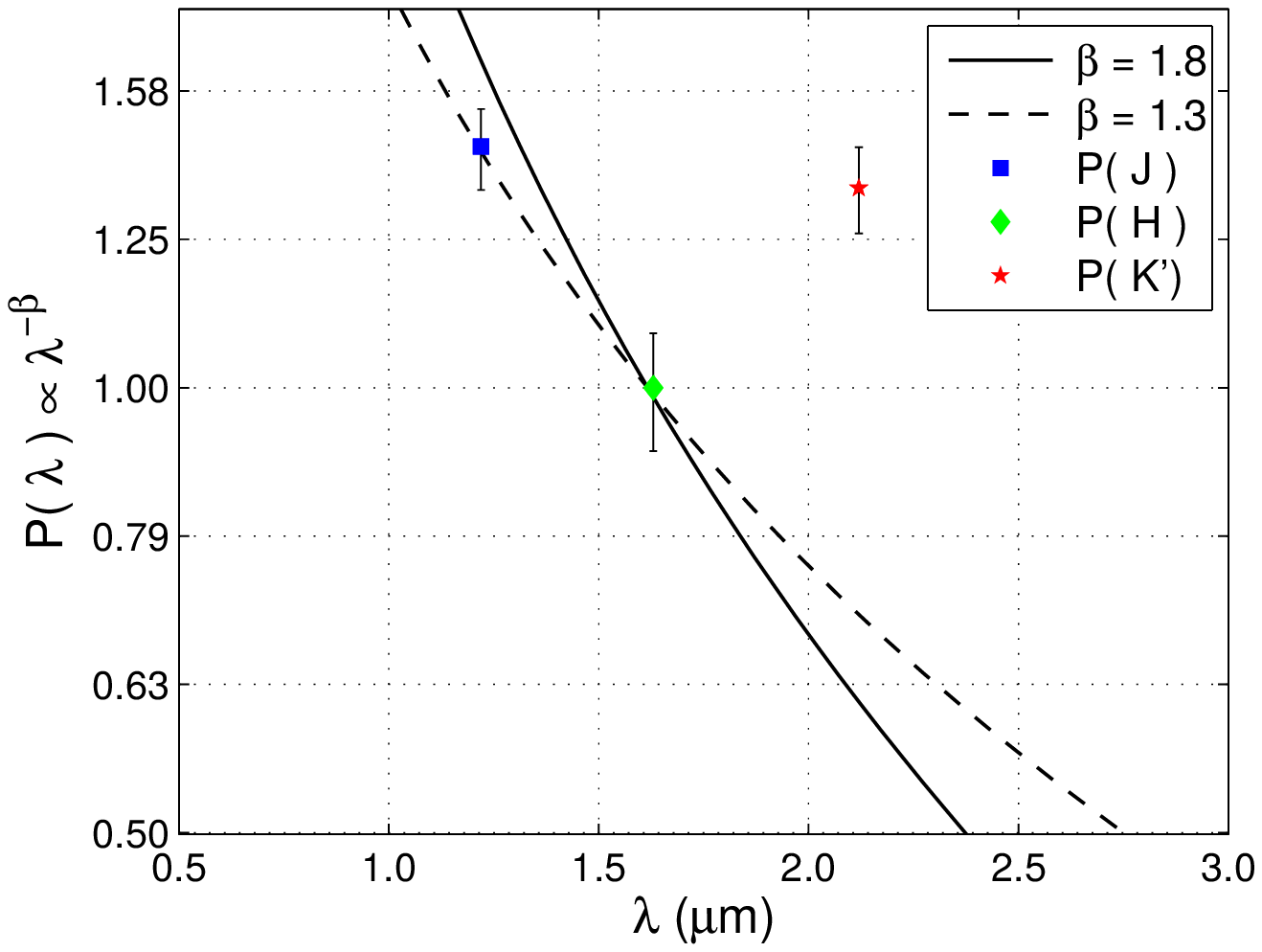}
    \caption{Average normalized polarization for the equatorial spot in J (square), H (diamond) and K' (star) bands.  
      A decreasing tendency between J and H bands is observed, an index of $\beta=1.3$ (dashed line) 
      is needed to fit a $P(\lambda)$ power-law. The ISM power-law with an index $\beta=1.8$ is 
      also plotted for comparison (solid line).
      The K' band polarization does not follow the power-law, this is probably because it 
      belongs to a deeper region (see text in sec. \ref{sec:magnetic}) and is plotted only to complete the NIR set of data.}
    \label{fig:power}
    \end{figure}
\subsection{Magnetic field nearby the central star of Frosty Leo?}
\label{sec:magnetic}
The polarization maps from \citet{Murakawa2008} show a spot of aligned vectors located at the equator in both, H and K bands.
The suggested mechanism of polarization in this area was the dichroism. A dichroic interpretation (directional extinction) requires aligned non-spherical grains.
Such grains should tend to rotate with their longer axes orientated perpendicularly to the angular momentum vector.
If a magnetic field is imposing the alignment, its direction should coincide with the angular momentum of the grains, thus,
the transmitted electric vector (the polarization) is parallel to the mean-field direction \citep{Lazarian2007,Whitney2002,Crutcher2012}.
Although our resolution is poorer (1\arcsec), our data
in J, H and K' bands also exhibit constant position angles in the same area (the equatorial spot ES) (see figs. \ref{fig:K} and \ref{fig:aligned}).
According to \citet{Li1997,Martin1992}, the dichroic ISM polarization at NIR wavelengths should follow the power-law 
$P(\lambda)\propto\lambda^{-\beta}$, where $\beta=1.8\pm 0.2$. This equation 
indicates that the polarization must follow a decreasing trend from shorter to larger wavelengths.
Therefore, if the NIR polarization of an object follows this trend,
it might belong to the dichroic case. If the coherent polarization seen in the ES is dichroic, 
this would indicate the presence of aligned grains and possibly, 
an ordering magnetic field with an orientation parallel to the equator. Otherwise, the multiple scattering plane proposed by \citet{Bastien1988}
would be a more suitable solution.\\
\begin{table}
\caption{Summary results of the polarization in Frosty Leo for regions.
         The labels in the column of Position Angles are C and A for 
         Centrosymmetric and Aligned, respectively.}
\label{tab:resumen}
 \begin{tabular}{cccccc}
 \hline
 Nebula's            &   J        &    H     &  K'       & Position             & Polarizing  \\
 region              & ($P\%$)    & ($P\%$)  &  ($P\%$)  & Angles ($^{\circ}$)  & mechanism   \\
 \hline
 Bipolar             &            &          &           &                      &  Single     \\
 Lobes               &    62      &   70     &    89     &       C              &  Scattering \\
 (max)               &            &          &           &                      &             \\
 \hline
 Extended            &            &          &           &                      &  Multiple   \\
 Envelope            &    16      &   20     &     -     &       A              & Scattering  \\
 (avg)               &            &          &           &                      &             \\  
 \hline
 Equatorial          &            &          &           &                      &             \\
 Spot                &    16      &   11     &    15     &       A              &  Dichroism? \\
 (avg)               &            &          &           &                      &             \\
 \hline
  \end{tabular}
\end{table}
We investigated if the polarization in Frosty Leo follow the $P(\lambda)$ power-law.  
To do so, we computed the average $P\%$ in the ES for each band and then 
compared them with the $P(\lambda)$ power-law for several values of $\beta$. 
We found a decreasing tendency in the average polarization between J and H bands (see table \ref{tab:resumen} and fig. \ref{fig:power}),
but the index required to make a fit is $\beta = 1.3$ which is smaller than the predicted by \citet{Li1997,Martin1992}.
The $P\%$ goes up in the K' band, also moving away from the curve of theoretical polarization.\\
Although not entirely conclusive, these results compel us to reconsider the dichroic interpretation in the ES. 
However, there still are some arguments that should be taken into account before discarding this possibility.
For example, the increase in K' band polarization could be explained by assuming that our map is probing deeper regions
than J and H. The K band data from \citet{Dougados1990} also supports this idea. 
If the light in our K' image is coming from a different layer, the polarization 
in this band might not be related to that in J and H or, caused by some other mechanism. Thus,
the aligned and decreasing $P\%$ between J and H bands, could still be a hint of a dichroic effect occurring 
in the outer layers.
Furthermore, recent submillimeter polarization studies of this source \citep{Sabin2019} also suggest
the presence of a magnetic field located at the equatorial region. The detection is marginal but the orientation of the 
inferred field would be parallel to the equator, just as the polarization in the NIR data suggests (\citealt{Murakawa2008} and our own data), 
so we should not hurry in dismissing this idea. Further 
polarization investigations at far infrared and millimeter wavelengths could help obtaining more insights into this subject.\\
\section{Conclusions}
 We studied the pre-planetary nebula Frosty Leo at NIR wavelengths using polarimetry and we detected the following features: 
\begin{itemize}
 \item The north and south shells (NS and SS) are seen in polarized emission. 
 These shells are located at the tips of the bipolar nebula.
 Their centrosymmetric vector pattern indicate that the light is being scattered. 
 They are clearly seen in J and H bands and cover large areas, in the K' band they are visible but less extended although 
 this is probably due to our detection limit. We propose that these shells might be the remains of a 
 fast ejection of material during a superwind event. The estimated duration is of 600 yr.
 \item The extended envelope (EE) which is seen in polarized light and total intensity. This halo seen in J and H bands, has a diameter 
 of 35\arcsec\ (in H band) and is the remaining material of the previous AGB phase. The PAs of the vectors associated to the EE
 indicate an alignment in J and H bands and are almost parallel to the equatorial plane of the nebula. 
 This behavior could be associated to multiple scattering since the light that reaches the EE is most likely coming
 from the bipolar lobes. Then, the material in the EE is probably scattering the light again into our line of sight resulting in the observed 
 polarization.
 \item We performed a test to determine if the polarization in the equatorial region is due to aligned grains.
 We found a decreasing polarization between J and H bands, but, it is not consistent to match the $P(\lambda)$ power-law 
 for the ISM. Further polarization research at larger wavelengths are needed
 to solve this issue.
\end{itemize}
\section*{Acknowledgements}
We would like to thank to CONACyT-M\'exico for the fundings assigned to the project CB A1S54450
and the scholarship with CVU 480840 to carry out this research. R.D. acknowledges CONACyT-M\'exico for SNI grant (CVU 555629)
and support from the European Research Council advanced Grant
H2020-ERC-2016-ADG-74302 under the European Union's Horizon 2020
Research and Innovation programme.
We thank to all the personnel from the Guillermo Haro Astrophysics Observatory  
for the valuable help they provided during the observations.
We also thank the anonymous referee for the valuable comments that helped 
to improve the quality of this paper.
This research made use of APLpy, an open-source plotting package for Python (Robitaille and Bressert, 2012).
%



\bibliographystyle{mnras}
\bibliography{frosty} 




\bsp	
\label{lastpage}
\end{document}